\documentclass[a4paper, amsfonts, amssymb, amsmath, reprint, showkeys, nofootinbib, twoside,notitlepage,onecolumn]{revtex4-1}

\def\apj{{ApJ}}                 

\def\prd{{Phys.~Rev.~D}}        
\def\prl{{Phys.~Rev.~Lett.}}    
\usepackage{amsmath,amstext}
\usepackage[T1]{fontenc}
\usepackage{amssymb}
\usepackage{graphicx}
\usepackage{ae,aecompl}

\DeclareFontFamily{OT1}{pzc}{}
\DeclareFontShape{OT1}{pzc}{m}{it}{<-> s * [1.10] pzcmi7t}{}
\DeclareMathAlphabet{\mathpzc}{OT1}{pzc}{m}{it}

\usepackage{hyperref}
\usepackage{amsmath}
\usepackage{amssymb}
\usepackage{mathtools}
\usepackage{bm}
\usepackage{cleveref}
\usepackage{tensor}
\usepackage{braket}
\usepackage{enumitem}
\usepackage{mhchem}
\usepackage{amsthm}
\usepackage{nccmath}
\usepackage{mathrsfs}
\usepackage{color}

\newcommand{\E}{{\mathbb{E}}}

\newcommand{\s}{{\sigma \!\! \! \sigma}}

\def\be{\begin{equation}}
\def\ee{\end{equation}}
\def\beq{\begin{eqnarray}}
\def\eeq{\end{eqnarray}}

\theoremstyle{definition}

\theoremstyle{theorem}

\theoremstyle{corollary}

\begin{document}
\title{Effective field theory of quasi-hydrodynamics from kinetic theory}
\author{L.~Gavassino}
\affiliation{Department of Applied Mathematics and Theoretical Physics, University of Cambridge, Wilberforce Road, Cambridge CB3 0WA, United Kingdom}

\begin{abstract}
Quasi-hydrodynamics describes systems with quasi-conserved degrees of freedom, namely observables that relax on timescales that are finite but parametrically longer than microscopic relaxation times. Examples include kinetic chemistry and linear viscoelasticity. Here, we develop a rigorous effective-field-theory framework for linear quasi-hydrodynamics from kinetic-type theories. Starting from any linearized, causal kinetic-like theory endowed with slow degrees of freedom, we show that the exact dynamics of conserved and quasi-conserved observables admits a systematic expansion in the fast relaxation timescale. At zeroth order, the resulting equations form a causal, symmetric-hyperbolic theory belonging to the appropriate transient-hydrodynamic universality class, establishing Israel-Stewart-like dynamics as the universal description of slow relaxation modes. Higher-order corrections can be computed systematically and inherit universal symmetry, Onsager, positivity, and causality constraints from the underlying microscopic theory.
\end{abstract} 
\maketitle

\vspace{-0.2cm}
\textbf{\textit{Introduction --}} In modern formulations, hydrodynamics is viewed as a classical Effective Field Theory (EFT) \cite{Glorioso2018}. This approach assumes that, at sufficiently long wavelengths and late times, the only relevant degrees of freedom of a many-body system are the conserved densities, whose dynamics is constrained by symmetries and conservation laws. The resulting equations of motion admit a universal structure that can be organized as a gradient expansion, with successive derivative corrections suppressed by powers of the mean free path \cite{Struchtrup2011Review}, which sets the EFT ultraviolet cutoff.

Recently, quasi-hydrodynamics, namely the extension of hydrodynamics that incorporates a small number of non-hydrodynamic modes whose relaxation times are parametrically longer than the microscopic collision time, has begun to undergo a similar systematization. Historically, quasi-hydrodynamic theories were often developed either as phenomenological frameworks (as in rheology \cite{MalkinIsayev2011,BAGGIOLI20201}, or Hydro+ \cite{StephanovHydroPlus:2017ghc}), or as a collection of largely disconnected formalisms (e.g. kinetic chemistry, viscoelasticity, and magnetohydrodynamics with Amp\`{e}re's law). However, in recent years, a different picture has begun to emerge \cite{BaggioliZaccone2022}. Insights from holography have shown that effective quasi-hydrodynamic descriptions can be systematically extracted from theories with weakly broken symmetries \cite{Grozdanov2019,Andrade:2019zey,AhnBaggioli:2025odk}. Moreover, it was shown that, when all degrees of freedom are even under the combined action of parity and time reversal, Onsager reciprocity implies that a broad family of phenomenological quasi-hydrodynamic theories can be systematically organized into universality classes described by symmetric-hyperbolic equations of motion \cite{GavassinoNonHydro2022,GavassinoSymmetric2022nff,GavassinoUniveraalityI2023odx}. These classes constitute different realizations of transient hydrodynamics \cite{Jou_Extended,Muller_book} (such as Cattaneo theory \cite{cattaneo1958} and Israel-Stewart theory \cite{Israel_Stewart_1979,Hishcock1983}), and are uniquely determined by the number and geometric character of the quasi-conserved degrees of freedom. This classification has, in turn, revealed deep connections between seemingly disparate theories \cite{PriouCOMPAR1991,GavassinoRadiazione,GavassinoGENERIC:2022isg,GavassinoKhalatnikov2022,GavassinoFarFromBulk:2023xkt,GavassinoUniversalityII:2023qwl,GavassinoBurgers:2023eoz,Hernandez:2025zxw}.

Taken together, these developments point to the existence of a common EFT description underlying a wide variety of quasi-hydrodynamic systems. A systematic microscopic derivation of this structure, however, is still lacking.

In this Letter, we establish a rigorous EFT framework for linearized quasi-hydrodynamics directly from kinetic-type theories. Given a finite set of slowly relaxing degrees of freedom coupled to microscopic modes with parametrically shorter relaxation times, we prove that the exact dynamics of the slow variables admit a systematic expansion in powers of the fast relaxation timescale, which serves as the ultraviolet cutoff of the EFT. At zeroth order, the equations reduce to a causal symmetric-hyperbolic theory belonging to the transient-hydrodynamic universality class uniquely determined by the number and tensorial character of the slow degrees of freedom (see the atlas in \cite{GavassinoUniveraalityI2023odx}), with familiar examples including the Cattaneo equation and Israel-Stewart theory. This establishes transient hydrodynamics as the universal leading-order EFT of quasi-hydrodynamic systems. This description remains accurate even when gradients are large on the scale set by the macroscopic relaxation times of the theory itself (but small compared to the 
cutoff). Higher-order terms appear as systematically computable gradient corrections, suppressed by the microscopic relaxation time and becoming important only near the ultraviolet cutoff scale. We illustrate the formalism with several explicit examples and derive universal bounds on first-order EFT contributions.

We adopt the metric signature $(-,+,+,+)$, and work in natural units: $c=\hbar=k_B=1$.

\textbf{\textit{Abstract kinetic-type framework --}} The starting point of our construction is the observation that kinetic theory, radiative transfer theory, and many other theories of matter can be cast into a common self-adjoint form \cite{GavassinoSymmetric2022nff,GavassinoUniveraalityI2023odx,RochaGavassinoFlucut:2024afv,GavassinoDisturbing:2026klp}.

Let $\Psi:\text{``Minkowski''}\to\mathcal H$ be a linearized field describing perturbations away from global equilibrium. Here, $\mathcal H$ is a complex Hilbert space encoding the microscopic degrees of freedom available at a fixed spacetime event, endowed with Onsager's inner product $(*,*)$. The latter is defined so that $\frac12\int_{\mathrm{space}}(\Psi,\Psi)d^3 x$
coincides with the quadratic free-energy perturbation associated with $\Psi$. For example, in the kinetic theory of a non-degenerate gas, the distribution function can be decomposed as $f=f_{\mathrm{eq}}+\sqrt{f_{\mathrm{eq}}}\,\Psi$, and one may take $\mathcal H=L^2(\mathcal M)$, where $\mathcal M$ is the manifold of allowed particle momenta. With this choice, the natural $L^2$ inner product coincides with Onsager's inner product \cite{DudynskiEkielJezewska1985,GavassinoGapless:2024rck,GavassinoConvergence:2024xwf,RochaGavassinoFlucut:2024afv,GavassinoDisturbing:2026klp}.

The crucial observation is that, when $\Psi$ is even under PT symmetry (a standard feature of kinetic theory \cite{Gavassino:2025PlasmaOscillations}), Onsager reciprocity and causality imply \cite{GavassinoSymmetric2022nff,GavassinoUniveraalityI2023odx,RochaGavassinoFlucut:2024afv,GavassinoDisturbing:2026klp} that the equations of motion take the universal form
\vspace{-0.2cm}
\begin{equation}\label{boltzmann}
\partial_t\Psi=-(\s+\E^j \partial_j)\Psi\, ,
\end{equation}
where $\s$ and $\E^j$ are self-adjoint operators on $\mathcal H$. The operator $\s$ describes relaxation towards local equilibrium (in kinetic theory, it is the collision integral) and is non-negative definite by stability, while the operators $\E^j$ describe transport (in kinetic theory, they are the velocity operators). Causality implies that the latter are bounded: for every unit vector $n_j$, one has $||n_j\E^j||\le 1$ (recall that $c=1$). In kinetic theory, this condition reduces to the familiar statement that particle velocities cannot exceed the speed of light. The EFT developed below is obtained by systematically expanding the exact dynamics generated by \eqref{boltzmann}.

\textbf{\textit{Separation of the spectrum --}} For plane-wave perturbations $\Psi\propto e^{ik_j x^j-i\omega t}$ (with $\omega,k_j\in\mathbb C$), equation \eqref{boltzmann} reduces to $(\s+ik_j\E^j)\Psi=i\omega\Psi$, which has the form of an eigenvalue problem. The relaxation rates $i\omega$ therefore coincide with the spectrum of the operator $\s+ik_j\E^j$. In particular, at $k_j=0$, the set of relaxation rates coincides with $\text{Spectrum}(\s)$, which is contained in $[0,+\infty]$, since $\s$ is self-adjoint and non-negative.

The defining assumption of quasi-hydrodynamics is the existence of a spectral separation between a slow and a fast relaxation sector. More precisely, we assume that there exist two timescales $\tau_S$ and $\tau_F\ll\tau_S$ such that $\text{Spectrum}(\s)\subset[0,1/\tau_S]\cup[1/\tau_F,+\infty]$, and that $\text{Spectrum}(\s)\cap[0,1/\tau_S]$ consists of finitely many eigenvalues of finite multiplicity. These eigenvalues are the \textit{quasi-hydrodynamic modes}, while $\text{Spectrum}(\s)\cap[1/\tau_F,+\infty]$ is the \textit{UV sector}. 

\begin{figure}[h!]
    \centering
\includegraphics[width=0.40\linewidth]{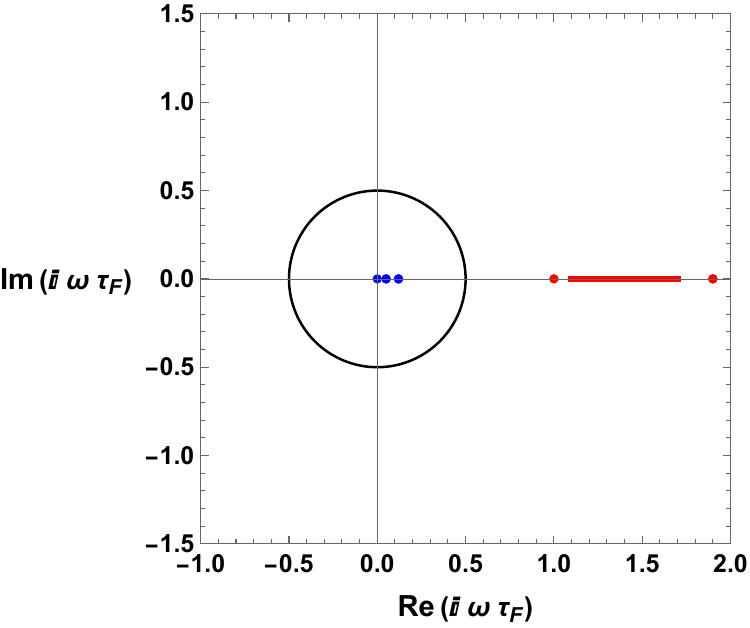}\hspace{0.08\linewidth}
\includegraphics[width=0.40\linewidth]{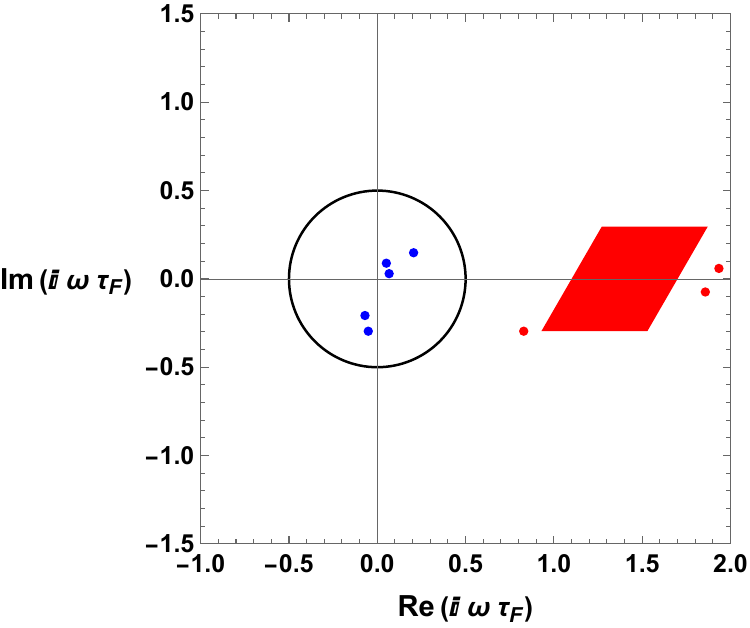}
\caption{Illustration of the quasi-hydrodynamic spectral separation. \textit{Left panel}: At $k_j=0$, the spectrum splits into a finite cluster of quasi-hydrodynamic modes (blue points) and a well-separated UV sector (red). The circle $\Gamma$, centered at the origin and of radius $(2\tau_F)^{-1}$, separates the two spectral components. \textit{Right panel}: Upon turning on a complex wavevector $k_j=kn_j$ (with $n_j\in S^2$ and $k\in\mathbb C$), the spectrum deforms continuously. Quasi-hydrodynamic eigenvalues may split and move into the complex plane, while continuous components of the UV spectrum may broaden into two-dimensional spectral branches. Nevertheless, as long as $|k|<k_{UV}$ [defined in equation \eqref{kuv}], Kato's perturbation theory guarantees that the circle $\Gamma$ continues to separate the quasi-hydrodynamic and UV sectors. Note that the estimate $k_{UV}\gtrsim (2\tau_F)^{-1}$ relies on the causal bound $||n_j\E^j||\le 1$.
}
    \label{fig:Circle}
\end{figure}

As a finite wavevector $k_j=kn_j$ (with $n_j\in S^2$ and $k\in\mathbb C$) is turned on, the quasi-hydrodynamic and UV sectors will in general move in the complex plane. However, the spectral separation survives for sufficiently small $k$. To estimate a minimal $k$-range, let $\Gamma$ be the circle in the complex $(i\omega)$-plane centered at the origin and with radius $1/(2\tau_F)$, which separates the two sectors (see figure \ref{fig:Circle}). The operators $\s+ik n_j\E^j$, viewed as functions of $k$, form a holomorphic family of type (A) in the sense of Kato \cite[\S VII.2]{Kato_Perturbation_Theory}. Hence, by \cite[\S VII.2.3, Remark 2.9]{Kato_Perturbation_Theory}, the quasi-hydrodynamic modes remain inside $\Gamma$, and the UV modes remain outside it, as long as $|k|<k_{UV}$ with
\vspace{-0.2cm}
\begin{equation}\label{kuv}
k_{UV}= \min_{z\in \Gamma} \dfrac{\operatorname{dist}\left(z,\operatorname{Spectrum}(\s)\right)}{||n_j\E^j||}\geq (2\tau_F)^{-1}-\tau_S^{-1}\approx (2\tau_F)^{-1} \, ,
\end{equation}
where we have used the causality condition $||n_j\E^j||\leq 1$. We can therefore identify $\tau_F$ as the microscopic scale that sets the ultraviolet cutoff of quasi-hydrodynamics.

\textbf{\textit{EFT of quasi-hydrodynamics --}} Taking $\Gamma$ to be the same circle as above (oriented counter-clockwise), we define the self-adjoint operators $\mathbb{S}=-\frac{1}{2\pi i} \oint_\Gamma (\s-z)^{-1} dz$ and $\mathbb{F}=1-\mathbb{S}$ \cite[\S III.6.4]{Kato_Perturbation_Theory}. These are orthogonal projectors ($\mathbb{S}=\mathbb{S}^2$, $\mathbb{F}=\mathbb{F}^2$, $\mathbb{S}\mathbb{F}=0$) \cite[\S V.3.5]{Kato_Perturbation_Theory}, which split the Hilbert space as $\mathcal{H}=\mathcal{H}_S\oplus \mathcal{H}_F$, where $\mathcal{H}_S$ (slow sector) is the Hilbert subspace generated by the quasi-hydrodynamic modes at $k_j=0$, while $\mathcal{H}_F=\mathcal{H}_S^\perp$ (fast sector) is the space generated by the UV modes at $k_j=0$.

Now, let $\Psi$ be a quasi-hydrodynamic mode associated with a wavevector $k_j=kn_j$ satisfying $|k|<k_{UV}$. Defining the decomposition $\Psi=\Psi_S+\Psi_F$, with $\Psi_S=\mathbb{S}\Psi\in \mathcal{H}_S$ and $\Psi_F=\mathbb{F}\Psi\in \mathcal{H}_F$, the eigenvalue equation $(\s+ik_j\E^j-i\omega)\Psi=0$ takes the block form
\vspace{-0.3cm}
\begin{equation}\label{breakingdown}
\begin{bmatrix}
\mathbb{S}(\s +ik_j\E^j-i\omega)\mathbb{S} & ik_j \mathbb{S}\E^j \mathbb{F} \\
ik_j \mathbb{F}\E^j \mathbb{S} &  \mathbb{F}(\s +ik_j\E^j-i\omega)\mathbb{F} \\
\end{bmatrix}
\begin{bmatrix}
\Psi_S\\
\Psi_F \\
\end{bmatrix}=0 \, .
\end{equation}
The compression $(\s_{\mathcal H_F}{+}ik_j{\E^j}_{\mathcal H_F}{-}i\omega)$ is invertible with bounded inverse\footnote{In operator theory, the compression $\mathbb{A}_{\mathcal{H}_F}:\mathcal{H}_F \to \mathcal{H}_F$ is simply the projected operator $\mathbb{F}\mathbb{A}\mathbb{F}$ viewed as an operator on $\mathcal{H}_F$. An analogous definition holds for $\mathbb{A}_{\mathcal{H}_S}$.}. In fact, $i\omega$ lies inside $\Gamma$, whereas
$\text{Spectrum}(\s_{\mathcal H_F}) \subseteq[1/\tau_F,+\infty]$, and 
$\text{Spectrum}(\s_{\mathcal H_F}{+}ik_j{\E^j}_{\mathcal H_F})$ lies within a distance
$\|ik_j{\E^j}_{\mathcal H_F}\|\le |k|$
of $\text{Spectrum}(\s_{\mathcal H_F})$
\cite[\S V.4.3, Problem 4.8]{Kato_Perturbation_Theory},
which implies that $\text{Spectrum}(\s_{\mathcal H_F}{+}ik_j{\E^j}_{\mathcal H_F})$ remains outside $\Gamma$ for $|k|<k_{UV}$.
Thus, we can solve the second line of \eqref{breakingdown}, obtaining
$\Psi_F=-ik_j(\s_{\mathcal H_F}+ik_l{\E^l}_{\mathcal H_F}-i\omega)^{-1}\mathbb{F}\E^j\mathbb{S}\Psi_S$. Substituting this expression into the first line yields a closed equation for the slow sector:
\vspace{-0.1cm}
\begin{equation}\label{exact}
(\s_{\mathcal H_S}+ik_j{\E^j}_{\mathcal H_S})\Psi_S
+k_jk_k\mathbb{S}\E^j \mathbb{F}
(\s_{\mathcal H_F}+ik_l{\E^l}_{\mathcal H_F}-i\omega)^{-1} \mathbb{F}
\E^k\mathbb{S}\Psi_S
=i\omega\Psi_S \, .
\end{equation}

Equation \eqref{exact} is exact. In the regime $|\omega|,|k|\ll\tau_F^{-1}$, one has $||(\s_{\mathcal H_F}+ik_l{\E^l}_{\mathcal H_F}-i\omega)^{-1}||\sim\tau_F$. Combined with the causal bound $||n_j\E^j||\le1$, this implies that the second term in \eqref{exact} is perturbative. We may therefore expand the resolvent through the Neumann series $[\s_{\mathcal H_F}-(i\omega-ik_l{\E^l}_{\mathcal H_F})]^{-1}=\sum_{a=0}^{\infty}[\s^{-1}_{\mathcal H_F}(i\omega-ik_l{\E^l}_{\mathcal H_F})]^a\s^{-1}_{\mathcal H_F}$, which is necessarily convergent for $|k|< (2\tau_F)^{-1}-\tau_S^{-1}$, since then $||\s^{-1}_{\mathcal H_F}(i\omega-ik_l{\E^l}_{\mathcal H_F})||\leq\tau_F (|\omega|+|k|)< 1$. Reconstructing a partial differential equation from the eigenvalue problem, we obtain the central result of this Letter:
\vspace{-0.2cm}
\begin{equation}\label{EFT}
\partial_t \Psi_S=-(\s_{\mathcal H_S}+{\E^j}_{\mathcal H_S}\partial_j)\Psi_S
+\mathbb{S}\E^j
\mathbb{F}\left[
\s^{-1}_{\mathcal H_F}
+\s^{-1}_{\mathcal H_F}(-\partial_t-{\E^l}_{\mathcal H_F}\partial_l)\s^{-1}_{\mathcal H_F}
+\cdots
\right]\mathbb{F}
\E^k\mathbb{S}\,
\partial_j\partial_k\Psi_S \, .
\end{equation}

As in every EFT, the equations of motion are organized as a derivative expansion. Unlike ordinary hydrodynamics, however, not all derivatives are counted as small. Only derivatives accompanied by a power of $\s^{-1}_{\mathcal H_F}$ are suppressed, since $||\s^{-1}_{\mathcal H_F}||\leq \tau_F$, which is the microscopic scale of the system. By contrast, the scale $\tau_S$ appearing in $\s_{\mathcal H_S}$ is treated as order unity in the power counting. This is the defining feature of the quasi-hydrodynamic EFT expansion.

Note that, if one truncates the EFT expansion at some finite order, the resulting equations of motion will in general contain time derivatives of order higher than one. The additional degrees of freedom associated with these higher derivatives are artifacts of the truncation and do not belong to the exact EFT. They may be systematically removed using the order-reduction procedure, as outlined in \cite{GavassinoReallInitialData:2026xjw}.

\textbf{\textit{Some explicit examples --}} We now compare two distinct UV theories that possess the same quasi-hydrodynamic degrees of freedom, but differ in their fast sectors, and hence in their higher-order EFT corrections. In both cases, we consider planar shear waves with gradients along the $x$ direction and velocity fluctuations $\delta u^j$ polarized along the $y$ direction. The slow non-hydrodynamic mode is a shear-stress contribution $\delta\pi^{xy}_S$, whose zeroth-order dynamics is governed by an Israel-Stewart relaxation equation \cite{Israel_Stewart_1979,Hishcock1983}. Many liquids are known to have a quasi-hydrodynamic behavior of this kind \cite[\S 36]{landau7}, and in rheology are classified as viscoelastic substances of Maxwell's type \cite{Trachenko2016,BAGGIOLI20201}.

\textit{Model 1 --} Our first UV theory is the Burgers model \cite{GavassinoUniversalityII:2023qwl} (or MIS$^*$ \cite{KeMISstar:2022tqf}), which splits the total shear stress as the sum of a fast and a slow contribution, both of which undergo Israel-Stewart-type relaxation. The equations of motion then read $(\varepsilon{+}P)\partial_t \delta u^y+\partial_x \delta\pi^{xy}_S+\partial_x \delta\pi^{xy}_F=0$, and $\tau_{S,F} \partial_t \delta\pi^{xy}_{S,F}+\delta\pi^{xy}_{S,F}=-\eta_{S,F}\partial_x \delta u^y$, where $\varepsilon{+}P$ is the enthalpy density, while $\tau_{S},\tau_F,\eta_S,\eta_F$ are transport coefficients. By taking $\Psi=[\sqrt{\varepsilon{+}P}\,\delta u^y,\sqrt{\tau_S/\eta_S}\,\delta\pi_S^{xy},\sqrt{\tau_F/\eta_F}\,\delta\pi_F^{xy}]^T$, we find \cite{GavassinoUniversalityII:2023qwl} that the Onsager inner product is just $(\Psi,\Phi)=\Psi^{T*}\Phi$, so that $\mathcal{H}=\mathbb{C}^3$. As a result, introducing the speeds $c_{S,F}=\sqrt{\eta_{S,F}/[(\varepsilon{+}P)\tau_{S,F}]}$, the equations of motion naturally take the form \eqref{boltzmann}:
\vspace{-0.2cm}
\begin{equation}
\begin{cases}
\partial_t \Psi_1 +c_S \partial_x \Psi_2 + c_F \partial_x \Psi_3 =0 \, , \\
\partial_t \Psi_2 +\tau_S^{-1}\Psi_2+c_S \partial_x \Psi_1 = 0 \, , \\
\partial_t \Psi_3 +\tau_F^{-1}\Psi_3+c_F \partial_x \Psi_1 = 0 \, , \\
\end{cases} \qquad \Longrightarrow \qquad \s+\E^x\partial_x =
\begin{bmatrix}
0 & 0 & 0 \\
0 & \tau_S^{-1} & 0 \\
0 & 0 & \tau_F^{-1} \\
\end{bmatrix}+
\begin{bmatrix}
0 & c_S & c_F \\
c_S & 0 & 0 \\
c_F & 0 & 0 \\
\end{bmatrix} \partial_x \, .
\end{equation}

The EFT arises in the limit where $\tau_F \to 0$. In this case, the slow and fast projectors are just $\mathbb{S}=\text{diag}(1,1,0)$ and $\mathbb{F}=\text{diag}(0,0,1)$, and the EFT \eqref{EFT} can be written explicitly up to infinite order:
\vspace{-0.3cm}
\begin{equation}\label{model1}
\partial_t 
\begin{bmatrix}
\Psi_1 \\
\Psi_2 \\
\end{bmatrix}= -
\begin{bmatrix}
0 & c_S \partial_x \\
c_S \partial_x & \tau_S^{-1}\\
\end{bmatrix}
\begin{bmatrix}
\Psi_1 \\
\Psi_2 \\
\end{bmatrix}+\tau_F c_F^2
\begin{bmatrix}
1 & 0 \\
0 & 0 \\
\end{bmatrix}
\sum_{a=0}^\infty (-\tau_F \partial_t)^a
\partial_x^2
\begin{bmatrix}
\Psi_1 \\
\Psi_2 \\
\end{bmatrix} \, .
\end{equation}
At zeroth order in $\tau_F$, we recover Israel-Stewart theory with only the slow stress $\delta\pi_S^{xy}$. At first order, the fast contribution enters as a Navier-Stokes term, $\delta\pi^{xy}_F=-\eta_F \partial_x \delta u^y$. Higher orders model the retarded response of $\delta\pi_F^{xy}$.

\textit{Model 2 --} Let us consider a case with $\dim(\mathcal{H})=\infty$. Take $\Psi=\{\Psi_1,\Psi_2,\psi(n^j)\}$, where $\Psi_1$ is proportional to the total density of momentum of the liquid in direction $y$, $\Psi_2$ is proportional to a slow shear stress component, and $\psi\in L^2(S^2)$ is some fast-relaxing angular distribution of quasi-particles. Suppose that the Onsager inner product is $(\Psi,\Phi)=\Psi_1^*\Phi_1+\Psi_2^*\Phi_2+\langle \psi^*\phi\rangle$, where $\langle * \rangle$ denotes the average over all angles. Then, $\mathcal{H}=\mathbb{C}^2\oplus L^2(S^2)$, and the following model equations fulfill all the structural requirements of \eqref{boltzmann}:
\vspace{-0.3cm}
\begin{equation}\label{eqmodel3}
\begin{cases}
\partial_t \Psi_1 +c_S \partial_x \Psi_2 +c_1\partial_x\langle \psi n^xn^y \rangle =0 \, , \\
\partial_t \Psi_2+\tau_S^{-1}\Psi_2 +c_S \partial_x \Psi_1 +c_2\partial_x\langle \psi n^y \rangle =0 \, , \\
\partial_t \psi+c_0 n^x\partial_x \psi+\tau_F^{-1}\psi+c_1 n^x n^y \partial_x\Psi_1 +c_2 n^y \partial_x\Psi_2=0 \, . \\
\end{cases}
\end{equation}
The first equation is the momentum conservation law $\partial_t \delta T^{0y}+\partial_x \delta T^{xy}=0$, where  $\delta T^{xy}$ carries contributions from both $\Psi_2$ and $\langle \psi n^x n^y\rangle$. The second equation is an Israel-Stewart-type relaxation equation for $\Psi_2$, where the source term $\partial_x \text{``flow in direction }y\text{''}$ carries contributions from $\partial_x \Psi_1$ and $\partial_x \langle \psi n^y \rangle$. The third line is the Boltzmann equation for the quasi-particles, where the last two terms are dictated by thermodynamics \cite{GavassinoSymmetric2022nff,GavassinoUniveraalityI2023odx,GavassinoDisturbing:2026klp}.

\begin{figure}[h!]
    \centering
\includegraphics[width=0.42\linewidth]{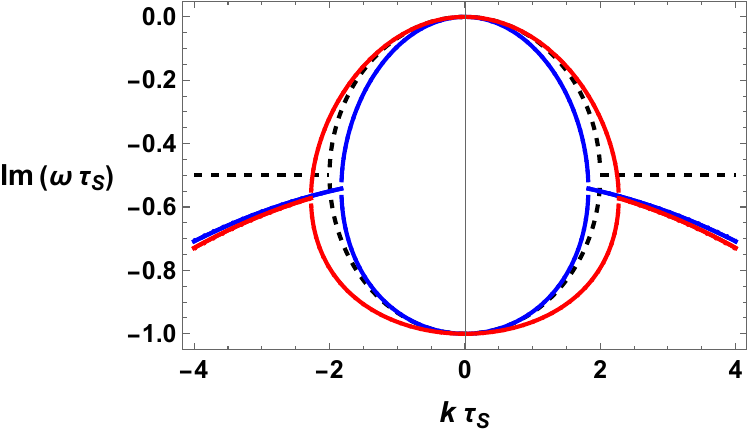}\hspace{0.08\linewidth}
\includegraphics[width=0.42\linewidth]{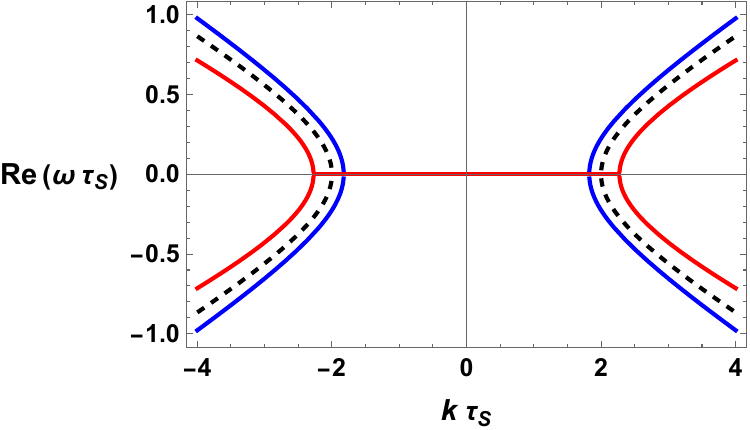}
\caption{Comparison between the Fourier modes of Model 1 (blue) and Model 2 (red) for $\{\tau_F/\tau_S,c_S,c_F\}=\{0.1,1/4,1/2\}$. Both models describe shear waves in a viscoelastic medium of Maxwell type, and at zeroth order reduce to Israel-Stewart theory of shear viscosity (black dashed). At higher orders, however, they behave differently. The plots show both the exact quasi-normal modes of the UV theories (dotted) and the EFTs \eqref{model1} and \eqref{model2} truncated at second order (continuous), which overlap perfectly.
}
    \label{fig:Models12}
\end{figure}

For small $\tau_F$, we obtain the following EFT:
\vspace{-0.3cm}
\begin{equation}\label{model2}
\partial_t 
\begin{bmatrix}
\Psi_1 \\
\Psi_2 \\
\end{bmatrix}= -
\begin{bmatrix}
0 & c_S \partial_x \\
c_S \partial_x & \tau_S^{-1}\\
\end{bmatrix}
\begin{bmatrix}
\Psi_1 \\
\Psi_2 \\
\end{bmatrix}+\tau_F \sum_{a=0}^\infty (-\tau_F)^a\langle 
\begin{bmatrix}
c_1^2 n_x^2 & c_1 c_2 n_x \\
c_1 c_2 n_x & c_2^2 \\
\end{bmatrix}
n_y^2 (\partial_t{+}c_0 n^x \partial_x)^a \rangle
\partial^2_x 
\begin{bmatrix}
\Psi_1 \\
\Psi_2 \\
\end{bmatrix}\, .
\end{equation} 
The Fourier modes of Models 1 and 2 are compared in figure \ref{fig:Models12}, for $c_0=c_1=0$ and $c_2=\sqrt{3} c_F$.

The examples above are simplified by the fact that $\s$ is diagonal from the outset and $\s_{\mathcal H_F}= \tau_F^{-1}$. In more realistic systems, the fast sector possesses a non-trivial internal structure, and the evaluation of the EFT expansion requires a more careful operator analysis. In the Supplementary Material, we illustrate the required techniques by deriving the zeroth- and first-order EFT contributions to shear waves in a viscoelastic liquid coupled to photon radiation.

\textbf{\textit{Classifying and constraining EFTs --}} The main advantage of the EFT framework is that many of its structural properties do not depend on the microscopic theory. Once the quasi-hydrodynamic degrees of freedom are identified, symmetry and general operator-theoretic arguments determine the most general form of the effective equations and constrain many of their coefficients. Let us see how this works.

Suppose that the background equilibrium state is isotropic. Then, fixed a rotation $R$, if $\Psi(t,\mathbf{x})$ is a solution of \eqref{boltzmann}, also the rotated process $\Psi_R(t,\mathbf{x})=\mathbb{U}(R)\Psi(t,R^{-1}\mathbf{x})$ is a solution of the same equation\footnote{Note that in Models 1 and 2 we have focused exclusively on planar shear waves with gradients along the $x$ direction and velocity fluctuations along the $y$ direction. Hence, the only relevant rotational symmetries of those models are $180^\circ$ rotations about the $x$, $y$, and $z$ axes.}. Here, $\mathbb{U}(R):\mathcal{H}\to\mathcal{H}$ is an invertible operator implementing the action of $R$ on $\mathcal{H}$. Since the free-energy density is a rotational scalar, the Onsager inner product is invariant under rotations, and therefore $\mathbb{U}(R)$ is unitary. Requiring that $\Psi_R(t,\mathbf{x})$ solve \eqref{boltzmann} for every solution $\Psi(t,\mathbf{x})$ implies that $\mathbb{U}(R)^\dagger\s\mathbb{U}(R)=\s$ and $\mathbb{U}(R)^\dagger\E^j\mathbb{U}(R)=R\indices{^j_k}\E^k$. In particular, $\mathbb{U}(R)^\dagger\mathbb{S}\mathbb{U}(R)=\mathbb{S}$, so $\mathcal{H}_S$ is an invariant subspace of $\mathbb{U}(R)$. Hence, $\mathcal{H}_S$ decomposes into irreducible unitary representations of the rotation group, and Schur's lemma implies that $\s$ is proportional to the identity within each irreducible representation. Thus, the slow degrees of freedom naturally split into irreducible rotational tensor structures (scalars, vectors, symmetric traceless tensors, ...), each of which is either conserved (if $\s=0$) or quasi-conserved (if $\s>0$).

The effective operators inherit the fundamental structural properties of the UV theory. In particular, $\s_{\mathcal H_S}$ and ${\E^j}_{\mathcal H_S}$ are self-adjoint (indeed, real symmetric, in all relevant cases), $\s_{\mathcal H_S}$ is non-negative, and $||n_j{\E^j}_{\mathcal H_S}||\le1$. Morevoer, $\mathbb{U}(R)^\dagger\s_{\mathcal H_S}\mathbb{U}(R)=\s_{\mathcal H_S}$ (i.e. $\s_{\mathcal H_S}$ is a rotational scalar), while $\mathbb{U}(R)^\dagger{\E^j}_{\mathcal H_S}\mathbb{U}(R)=R\indices{^j_k}{\E^k}_{\mathcal H_S}$ (i.e. ${\E^j}_{\mathcal H_S}$ is a rotational vector). These properties combined imply that the zeroth-order EFT is a causal symmetric-hyperbolic theory of the kind discussed in \cite{GavassinoUniveraalityI2023odx}. Hence, it belongs to one of the transient-hydrodynamic universality classes, and can be systematically constructed by writing down the most general information current compatible with the tensorial content of the slow degrees of freedom (see \cite{GavassinoUniveraalityI2023odx,GavassinoUniversalityII:2023qwl} for more details). In particular, the fact that Models 1 and 2 are governed by Israel-Stewart theory at zeroth order is not just a consequence of their simplicity. Rather, it follows from the tensorial content of their slow sector: any shear channel with a conserved vector and a quasi-conserved symmetric traceless tensor necessarily belongs to the Israel-Stewart universality class.

A similar reasoning constrains the higher-order EFT corrections. In fact, equation \eqref{EFT} takes the general form
\begin{equation}\label{EFTBlind}
\partial_t \Psi_S=-(\s_{\mathcal H_S}+{\E^j}_{\mathcal H_S}\partial_j)\Psi_S+\tau_F\mathbb{D}^{jk}\partial_j \partial_k \Psi_S-\tau_F^2\mathbb{B}^{j\mu k}\partial_\mu \partial_j \partial_k \Psi_S +... \, ,
\end{equation}
and one finds that the EFT operators transform as rotational tensors, e.g. $\mathbb{U}(R)^\dagger\mathbb{D}^{jk}\mathbb{U}(R)=\mathbb{D}^{lh}R\indices{^j_l}R\indices{^k_h}$. Moreover, they inherit Onsager reciprocity from the microscopic theory, e.g. $[\mathbb{D}^{jk}]^\dagger\,{=}\,\mathbb{D}^{jk}$, where we further impose $\mathbb{D}^{jk}\,{=}\,\mathbb{D}^{kj}$, because $\partial_j\partial_k$ is symmetric. Equation \eqref{EFT} also implies positivity constraints on the EFT coefficients, such as $(\Phi_{(j)},\mathbb{D}^{jk}\Phi_{(k)})\ge0$ for any triplet of vectors $\Phi_{(j)}\in\mathcal H_S$, together with causality bounds such as $||n_j\mathbb{D}^{jk}n_k||\le1$. Note that the positivity of $\mathbb{D}^{jk}$ implies that first-order corrections always increase dissipation, since
\begin{equation}
\dfrac{d}{dt}\underbrace{\int d^3x\,\dfrac{1}{2}(\Psi_S,\Psi_S)}_{\text{Free energy of the slow variables}}
=
-\int d^3x\,
\bigg[
\underbrace{(\Psi_S,\s_{\mathcal H_S}\Psi_S)}_{\ge0}
+\tau_F\underbrace{(\partial_j\Psi_S,\mathbb{D}^{jk}\partial_k\Psi_S)}_{\ge0}
\bigg]
+\mathcal O(\tau_F^2)\, .
\end{equation}
In summary, the EFT is constrained by symmetry, Onsager reciprocity, positivity, and causality. The most general quasi-hydrodynamic EFT is then obtained by writing down all tensor structures compatible with these requirements.

\textbf{\textit{Example --}} Suppose that the slow sector consists of a conserved rotational scalar $\rho$ and a quasi-conserved rotational vector $J^j$. Then, the resulting EFT belongs to the Cattaneo universality class \cite{GavassinoUniveraalityI2023odx}. Assuming that the variables have been normalized so that $(\Psi_S,\Psi_S)=|\rho|^2+J_j^*J^j$, the most general equations of motion up to first order in the EFT expansion take the form
\begin{equation}
\begin{split}
& \partial_t \rho + c_S \partial_j J^j = \tau_F \mathfrak{D}_1 \partial_j \partial^j \rho+\mathcal{O}(\tau_F^2) \, ,\\
& \partial_t J_l +\tau_S^{-1} J_l +c_S \partial_l \rho=\tau_F \mathfrak{D}_2 \partial^j  \left(\partial_j J_l+\partial_l J_j-\dfrac{2}{3}\delta_{jl}\partial_k J^k\right)+\tau_F \mathfrak{D}_3 \partial_l \partial_k J^k +\mathcal{O}(\tau_F^2) \, ,
\end{split}
\end{equation}
where $|c_S|\leq1$, $0\leq\mathfrak{D}_{n}\leq1$, and $\frac{4}{3}\mathfrak{D}_2+\mathfrak{D}_3\leq1$. In the Supplementary Material, we derive these equations from an explicit kinetic theory and compute all the EFT coefficients.



\textbf{\textit{Conclusions --}} We have established a rigorous EFT framework for quasi-hydrodynamics directly from kinetic-type theories. Starting from a general microscopic evolution equation satisfying only Onsager reciprocity, dissipation, and causality, we proved that the dynamics of the slow degrees of freedom admits a systematic expansion in the fast relaxation timescale. The construction is fully controlled at small wavenumbers: it follows from a convergent Neumann expansion of an exact equation for the slow sector, providing rigorous control over the truncation error together with a systematic procedure to compute corrections to arbitrary order.

Our results place transient hydrodynamics on rigorous microscopic foundations. At zeroth order, the EFT is always a causal symmetric-hyperbolic theory belonging to the transient-hydrodynamic universality class determined by the number and tensorial character of the slow degrees of freedom, as in \cite{GavassinoUniveraalityI2023odx}. Higher-order terms quantify systematic departures from transient hydrodynamics while inheriting symmetry, Onsager reciprocity, and positivity constraints from the underlying microscopic theory. Microscopic causality plays a central role, as it is needed to guarantee the persistence of the spectral separation between slow and fast modes, and therefore the convergence of the EFT expansion itself. Although finite-order truncations beyond zeroth order are generically acausal as systems of partial differential equations, BDNK-type reformulations can restore causality order by order without modifying the physical content of the EFT (we demonstrate this explicitly for the first-order theory in the Supplementary Material).

Many relativistic systems naturally possess a quasi-hydrodynamic structure. Examples include neutron-star matter (where weak interactions control slow bulk relaxation while electromagnetic interactions provide a much faster microscopic sector \cite{AlfordRezzolla}), Hydro+ \cite{StephanovHydroPlus:2017ghc}, radiation hydrodynamics, kinetic chemistry, and relativistic rheology. The present framework provides a systematic route to deriving, constraining, and extending quasi-hydrodynamic EFTs across these diverse settings.

\section*{Acknowledgements}

This work is supported by a MERAC Foundation prize grant,  an Isaac Newton Trust Grant, and funding from the Cambridge Centre for Theoretical Cosmology.

\bibliography{Biblio}

\newpage

\onecolumngrid
\newpage
\begin{center}
\textbf{\large Effective field theory of quasi-hydrodynamics from kinetic theory\\Supplementary Material}\\[.2cm]
  L. Gavassino\\[.1cm]
  {\itshape Department of Applied Mathematics and Theoretical Physics, University of Cambridge, Wilberforce Road, Cambridge CB3 0WA, United Kingdom\\}
\end{center}

\setcounter{equation}{0}
\setcounter{figure}{0}
\setcounter{table}{0}
\setcounter{page}{1}
\renewcommand{\theequation}{S\arabic{equation}}
\renewcommand{\thefigure}{S\arabic{figure}}

\section{Radiative corrections to Maxwell's viscoelasticity}

In this section, we illustrate, through a concrete example, the step-by-step construction of a quasi-hydrodynamic EFT from a realistic kinetic-like UV theory.

\subsection*{\textit{Step 1:} Identify the UV theory}

Consider shear perturbations in a viscoelastic liquid $L$ of Maxwell type, by which we mean a substance that, when taken in isolation, obeys Israel--Stewart theory, namely $(\varepsilon_L+P_L)\partial_t\delta u^y+\partial_x\delta\pi_S^{xy}=0$ and $\tau_S\partial_t\delta\pi_S^{xy}+\delta\pi_S^{xy}=-\eta_S\partial_x\delta u^y$, even at frequencies much larger than $\tau_S^{-1}$. Let this liquid be coupled to a radiation gas, whose perturbations are described by a kinetic distribution function $\delta f(x^\mu,p^\alpha)$. The photons are taken to have a mean free time $\tau_F$ satisfying $\tau_F\ll\tau_S$, but still much larger than the microscopic timescale at which the Israel--Stewart description of the liquid itself breaks down. Then, assuming that photons are absorbed and emitted according to the Kirchhoff--Planck law\footnote{Strictly speaking, the Kirchhoff--Planck law presupposes that the liquid is in local thermodynamic equilibrium, which is clearly not the case when $\delta\pi_S^{xy}\neq0$. In the present analysis, however, we will neglect for simplicity all shear-tensor-induced corrections to the emission and absorption rates.}, the dynamics is governed by the following equations of motion \cite{GavassinoNonNewtonianRadiation:2024jej}:
\begin{equation}\label{modelradiation}
\begin{split}
& (\varepsilon_L{+}P_L)\partial_t \delta u^y +\partial_x \delta\pi^{xy}_S +\int \dfrac{2d^3 p}{(2\pi)^3} p^y\left(\partial_t + \dfrac{p^x}{p^t}\partial_x \right) \delta f=0 \, , \\
& \tau_S \partial_t \delta\pi_S^{xy}+\delta\pi^{xy}_S=-\eta_S \partial_x \delta u^y \, , \\
& \left(\partial_t + \dfrac{p^x}{p^t}\partial_x \right) \delta f=\dfrac{1}{\tau_F} \left[f_{BB}(1{+}f_{BB})\beta p^y \delta u^y-\delta f \right] \, . \\
\end{split}  
\end{equation}
The first equation is the momentum conservation law for the total ``liquid$+$radiation'' system, and the third equation is the Boltzmann equation for photons (with $f_{BB}=(e^{p^t/T}{-}1)^{-1}$ the black-body distribution) under the gray opacity assumption, so that $\tau_F$ is a constant.

The gray opacity assumption is particularly useful for calculations because, defined 
$n^j=p^j/p^t$, we can consistently assume that the dependence of $\delta f$ on the energy $p^t$ and the photon direction $n^j$ factorize as follows: $\delta f=f_{BB}(1{+}f_{BB})\beta p^t \delta I(n^j)$, where $\delta I$ is an energy-independent angular distribution. Then, \eqref{modelradiation} simplifies to 
\begin{equation}\label{modelwithI}
\begin{split}
& (\varepsilon_L{+}P_L)\partial_t \delta u^y +\partial_x \delta\pi^{xy}_S +4aT^4 \tau_F^{-1} \langle n^y (n^y \delta u^y -  \delta I)\rangle=0 \, , \\
&  \partial_t \delta\pi_S^{xy}+ \tau_S^{-1}\delta\pi^{xy}_S+\dfrac{\eta_S}{\tau_S} \partial_x \delta u^y=0 \, , \\
& \partial_t \delta I + n^x\partial_x \delta I +\tau_F^{-1}(\delta I - n^y \delta u^y) =0 \, , \\
\end{split}  
\end{equation}
where $a\,{=}\,\pi^2/15$ is the radiation constant, and $\langle A(n^j)\rangle{=}\int_{S^2} A(n^j)\frac{d^2\Omega}{4\pi}$ is the angle average. We also used $\langle (n^y)^2\rangle \,{=}\,1/3$.

In the following, we will regard \eqref{modelwithI} as our UV theory.

\subsection*{\textit{Step 2:} Recast the UV theory in standard self-adjoint form}

The quadratic density of free energy associated with a perturbation $\{\delta u^y,\delta\pi_S^{xy},\delta I\}$ is just the sum of the free-energy densities of the liquid and the photon gas treated separately. Both have been computed in the past using information-current techniques \cite{GavassinoCausality2021,GavassinoNonHydro2022,RochaGavassinoFlucut:2024afv}. The final result is
\begin{equation}
\begin{split}
2\Delta \mathcal{F}=&  (\varepsilon_L{+}P_L)(\delta u^y)^2+\dfrac{\tau_S}{\eta_S} (\delta \pi^{xy}_S)^2 +4aT^4 \langle(\delta I)^2\rangle \, .\\
\end{split}
\end{equation}
This suggests that we define the degree of freedom $\Psi$ as follows:
\begin{equation}
\Psi=
\begin{bmatrix}
   \Psi_u \\
   \Psi_\pi \\
   \psi \\
\end{bmatrix}=
\begin{bmatrix}
 \sqrt{\varepsilon_L{+}P_L} \, \delta u^y\\
 \sqrt{\tau_S/\eta_S} \,\delta\pi_S^{xy}\\
 \sqrt{4aT^4} \, \delta I \\
\end{bmatrix} \, .
\end{equation}
In this way, the Onsager inner product is just $(\Psi,\Phi)=\Psi_u^*\Phi_u+\Psi_\pi^*\Phi_\pi+\langle \psi^*\phi \rangle$, so that $\mathcal{H}=\mathbb{C}^2 \oplus L^2(S^2)$, as in Model 2 of the main text.

Defined the elastic-wave speed
$c_L=\sqrt{\eta_S/[(\varepsilon_L{+}P_L)\tau_S]}$ of the liquid component, and the radiation dominance parameter $\mathcal{R}=\sqrt{4aT^4/(\varepsilon_L+P_L)}$, the system \eqref{modelwithI} becomes
\begin{equation}\label{systemselfadjradiation}
\begin{split}
& \partial_t \Psi_u +c_L\partial_x \Psi_\pi - \tau_F^{-1} \mathcal{R} \langle n^y(\psi-\mathcal{R}n^y \Psi_u)\rangle=0 \, , \\
&  \partial_t \Psi_\pi+ \tau_S^{-1}\Psi_\pi+c_L \partial_x \Psi_u=0 \, , \\
& \partial_t \psi + n^x\partial_x \psi +\tau_F^{-1}(\psi - \mathcal{R} n^y \Psi_u) =0 \, . \\
\end{split}  
\end{equation}
This has the standard form $\partial_t\Psi+\s\Psi+\E\partial_x\Psi=0$, with
\begin{equation}
\s
\begin{bmatrix}
\Psi_u \\
\Psi_\pi \\
\psi \\
\end{bmatrix}=
\begin{bmatrix}
 -\tau_F^{-1} \mathcal{R} \langle n^y(\psi-\mathcal{R}n^y \Psi_u)\rangle \\
\tau_S^{-1}\Psi_\pi \\
\tau_F^{-1}(\psi - \mathcal{R} n^y \Psi_u) \\
\end{bmatrix} \, , \qquad \qquad
\E
\begin{bmatrix}
\Psi_u \\
\Psi_\pi \\
\psi \\
\end{bmatrix}=
\begin{bmatrix}
c_L \Psi_\pi\\
c_L \Psi_u \\
n^x \psi \\
\end{bmatrix} \, .
\end{equation}
It is easy to verify that the above operators are indeed self-adjoint\footnote{In infinite dimensions, the symmetry property \((\Psi,\E\Phi)=(\E\Psi,\Phi)\) is not, by itself, sufficient to conclude that \(\E\) is self-adjoint, since the domains of \(\E\) and \(\E^\dagger\) may fail to coincide. For kinetic-type theories, however, this subtlety is absent. Indeed, causality implies that \(\E\) is bounded, while stability implies that \(\s\) is non-negative. These conditions, together with symmetry, guarantee that $\E$ and $\s$ always admit a self-adjoint extension. 
}, and fulfill the required properties, i.e. $(\Psi,\s\Psi)\geq 0$ and $(\Psi,\E\Psi)\in (\Psi,\Psi)[-1,1]$.

\subsection*{\textit{Step 3:} Identify the slow and fast Hilbert subspaces}

The decomposition $\mathcal{H}=\mathcal{H}_S\oplus \mathcal{H}_F$ depends on a separation of the spectrum of $\s$. Hence, we must first diagonalize this operator. This is immediately achieved via the following orthogonal decomposition:
\begin{equation}
\Psi= \underbrace{\dfrac{\Psi_1}{\sqrt{1+\mathcal{R}^2/3}} \begin{bmatrix}
1 \\
0 \\
\mathcal{R}n^y\\
\end{bmatrix}}_{\Psi_1 e_1}  + \underbrace{\Psi_2 \begin{bmatrix}
0 \\
1 \\
0\\
\end{bmatrix}}_{\Psi_2 e_2} + \underbrace{\dfrac{\Psi_3}{\sqrt{1/3+\mathcal{R}^2/9}}  \begin{bmatrix}
-\mathcal{R}/3 \\
0 \\
n^y \\
\end{bmatrix}}_{\Psi_3 e_3} +\begin{bmatrix}
0 \\
0 \\
\psi^\perp
\end{bmatrix} \qquad (\text{with }\psi^\perp \perp n^y) \,.
\end{equation}
In fact, we have $\s e_1=0$, $\s e_2 =\tau_S^{-1}e_2$, $\s e_3=\tau_F^{-1}(1+\mathcal{R}^2/3)e_3$, and $\s [0,0,\psi^\perp]^T=\tau_F^{-1} [0,0,\psi^\perp]^T$. Hence, we immediately conclude that
\begin{equation}
\mathcal{H}_S=\text{span}\{e_1,e_2\} \, , \qquad \qquad \mathcal{H}_F=\text{span} \{e_3, [0,0,\psi^\perp]^T\} \, .
\end{equation}
Since $(e_m,e_n)=\delta_{mn}$ (i.e., $e_n$ are orthonormal), the projector onto the slow subspace is $\mathbb{S}=e_1 (e_1,*)+e_2(e_2,*)$, and thus $\Psi_S=\Psi_1 e_1+\Psi_2 e_2$.

\subsection*{\textit{Step 4:} Zeroth-order theory}

Given that $\{e_1,e_2\}$ are an orthonormal basis of $\mathcal{H}_S$, the compressions $\s_{\mathcal{H}_S}$ and $\E_{\mathcal{H}_S}$ are represented by $2\times 2$ matrices whose entries are the corresponding matrix elements of $\s$ and $\E$:
\begin{equation}
\begin{split}
\s_{\mathcal{H}_S}=&
\begin{bmatrix}
(e_1,\s e_1) & (e_1,\s e_2) \\
(e_2,\s e_1) & (e_2,\s e_2) \\
\end{bmatrix}=
\begin{bmatrix}
0 & 0 \\
0 & \tau_S^{-1} \\
\end{bmatrix} \, , \\ 
\E_{\mathcal{H}_S}=&
\begin{bmatrix}
(e_1,\E e_1) & (e_1,\E e_2) \\
(e_2,\E e_1) & (e_2,\E e_2) \\
\end{bmatrix}= 
\begin{bmatrix}
0 & c_S \\
c_S & 0 \\
\end{bmatrix}  \qquad\qquad \text{with } \, \, c_S=\dfrac{c_L}{\sqrt{1+\mathcal{R}^2/3}} \, .
\end{split}
\end{equation}
The value of the effective speed $c_S$ has a very intuitive physical interpretation. In fact, it is the same as the elastic speed of the liquid component alone, $c_L=\sqrt{\eta_S/[(\varepsilon_L{+}P_L)\tau_S]}$, where the enthalpy density $\varepsilon_L{+}P_L$ of the liquid is now replaced by the conglomerate enthalpy density $\varepsilon_L{+}P_L{+}\frac{4}{3}aT^4$ of the total ``liquid$+$radiation'' system. This tells us that, in the limit where $\tau_F\to 0$, the photon gas participates in shear waves as an additional inertial component that comoves with the liquid. It does not modify the transport coefficients, but it enters the equation of state as an additive contribution to energy density and pressure.

\vspace{-0.2cm}
\subsection*{\textit{Step 5:} EFT corrections}
\vspace{-0.2cm}

To first order, we have $\partial_t \Psi_S=-(\s_{\mathcal{H}_S}+\E_{\mathcal{H}_S}\partial_x)\Psi_S+\mathbb{D}\partial^2_x\Psi+\mathcal{O}(\tau_F^2)$, with $\mathbb{D}=\mathbb{S}\E\mathbb{F}\s_{\mathcal{H}_F}^{-1}\mathbb{F}\E \mathbb{S}$, viewed as an operator on $\mathcal{H}_S$. Again, since we are working in the orthonormal basis $\{e_1,e_2\}$, we can write
\begin{equation}
\mathbb{D}=
\begin{bmatrix}
(e_1,\mathbb{S}\E\mathbb{F}\s_{\mathcal{H}_F}^{-1}\mathbb{F}\E \mathbb{S}e_1) & (e_1, \mathbb{S}\E\mathbb{F}\s_{\mathcal{H}_F}^{-1}\mathbb{F}\E \mathbb{S}e_2) \\
(e_2, \mathbb{S}\E\mathbb{F}\s_{\mathcal{H}_F}^{-1}\mathbb{F}\E \mathbb{S}e_1) & (e_2,\mathbb{S}\E\mathbb{F}\s_{\mathcal{H}_F}^{-1}\mathbb{F}\E \mathbb{S}e_2) \\
\end{bmatrix} \, .
\end{equation}
Thus, computing EFT corrections amount to computing matrix elements. Keeping in mind that $\s_{\mathcal{H}_F}^{-1}e_3{=}\tau_F(1{+}\mathcal{R}^2/3)^{-1}e_3$ and $\s_{\mathcal{H}_F}^{-1} [0,0,\psi^\perp]^T=\tau_F [0,0,\psi^\perp]^T$, we obtain
\begin{equation}
\begin{split}
& \mathbb{F}\E \mathbb{S}e_1= \dfrac{\mathcal{R}}{\sqrt{1+\mathcal{R}^2/3}} \begin{bmatrix}
0 \\
0 \\
n^x n^y\\
\end{bmatrix} \qquad \Longrightarrow \qquad (e_1,\mathbb{D}e_1)= \dfrac{\mathcal{R}^2 \tau_F}{15(1+\mathcal{R}^2/3)} \, ,  \\
&\mathbb{F}\E \mathbb{S} e_2=- \dfrac{c_L \mathcal{R}}{1+\mathcal{R}^2/3} 
\begin{bmatrix}
-\mathcal{R}/3 \\
0 \\
n^y \\
\end{bmatrix} \qquad \Longrightarrow \qquad (e_2,\mathbb{D}e_2)=
 \dfrac{c_L^2 \mathcal{R}^2 \tau_F}{3(1+\mathcal{R}^2/3)^2} \, , \\
\end{split}
\end{equation}
while the off-diagonal terms vanish. 
Putting everything together, we obtain
\begin{equation}\label{EFTRadiation}
\partial_t 
\begin{bmatrix}
\Psi_1 \\
\Psi_2 \\
\end{bmatrix}
= -
\begin{bmatrix}
0 & c_S \partial_x \\
c_S \partial_x & \tau_S^{-1} \\
\end{bmatrix}
\begin{bmatrix}
\Psi_1 \\
\Psi_2 \\
\end{bmatrix}+ \dfrac{\mathcal{R}^2 \tau_F}{1{+}\mathcal{R}^2/3}
\begin{bmatrix}
1/15 & 0 \\
0 & c_S^2/3 \\
\end{bmatrix}
\partial^2_x \begin{bmatrix}
\Psi_1 \\
\Psi_2 \\
\end{bmatrix}+\mathcal{O}(\tau_F^2) \, .
\end{equation}
Higher-order corrections can be calculated in a similar fashion.

Note that, physically, the coefficient $\mathfrak{D}=(e_1,\mathbb{D}e_1)$ can be interpreted as the contribution to the shear diffusivity due to non-equilibrium photons. Recalling that the inertia of the fluid is determined by the total enthalpy density, we recover Weinberg's expression \cite{Weinberg1971} for the radiative shear viscosity:
\begin{equation}
\eta_\text{rad}=\left(\varepsilon_L{+}P_L{+}\frac{4}{3}aT^4\right)\mathfrak{D}= \dfrac{4}{15} aT^4 \tau_F \, .
\end{equation}

\vspace{-0.2cm}
\subsection*{Consistency check}
\vspace{-0.2cm}

As a last check, let us verify that the dispersion relations one obtains from the EFT coincide with those that one obtains by directly solving the UV theory, and truncating the result to first order in $\tau_F$.

To that end, consider again system \eqref{systemselfadjradiation}. Assuming a spacetime dependence of the form $\Psi \propto e^{ikx-i\omega t}$, we can solve for $\Psi_\pi$ and $\psi$ in terms of $\Psi_u$. Plugging the result into the equation of motion for $\Psi_u$, we obtain an equation that is linear $\Psi_u$. Canceling $\Psi_u$, we are then left with an \textit{exact} implicit function linking $\omega$ and $k$, namely
\begin{equation}\label{Modesselfadjradiation}
-i\omega  +\dfrac{\tau_S c_L^2 k^2 }{1-i\omega\tau_S}  + \dfrac{\mathcal{R}^2}{3\tau_F} \left[1 -3\bigg\langle \dfrac{(n^y)^2}{1-\tau_F(i\omega  - n^x ik)} \bigg\rangle \right]=0 \, . 
\end{equation}
Expanding in geometric series the denominator in the angle bracket, and truncating to first order in $\tau_F$, we obtain
\begin{equation}\label{Modesselfadjradiation2}
-i\omega  +\dfrac{\tau_S c_S^2 k^2 }{1-i\omega\tau_S}  + \dfrac{\mathcal{R}^2 \tau_F}{1+\mathcal{R}^2/3} \left[\dfrac{k^2}{15}+\dfrac{\omega^2}{3}  \right]+\mathcal{O}(\tau_F^2)=0 \, . 
\end{equation}
Take now the EFT \eqref{EFTRadiation}. Its dispersion relations fulfill the following equation:
\begin{equation}
-i\omega  +\dfrac{\tau_S c_S^2 k^2 }{1-i\omega\tau_S}  + \dfrac{\mathcal{R}^2 \tau_F}{1+\mathcal{R}^2/3} \left[\dfrac{k^2}{15}-\dfrac{i\omega}{3} \dfrac{\tau_S c_S^2 k^2}{1-i\omega\tau_S}  \right]+\mathcal{O}(\tau_F^2)=0 \, ,
\end{equation}
which is identical to \eqref{Modesselfadjradiation2}, except for the second term in the square bracket. Luckily, this difference is only apparent. In fact, we observe that $i\omega=\tau_S c_S^2 k^2/(1-i\omega\tau_S)+\mathcal{O}(\tau_F)$, meaning that the square brackets differ by a higher order term. This confirms the consistency of the expansion.

\section{A kinetic UV completion of Cattaneo's theory}

Here, we provide an example of a kinetic theory that possesses a quasi-hydrodynamic sector which, at zeroth order, is governed by Cattaneo's theory of heat conduction. For clarity, we will follow the same step-by-step procedure as in the previous section.

\subsection*{\textit{Step 1:} Identify the UV theory}

We consider an ideal gas of massless bosons characterized by a kinetic distribution function $f(x^\mu,p^\alpha)$. These particles collide with each other over a short timescale $\Tilde{\tau}_F$ (the reason for the ``tilde'' will become clear later), and they can also be created and annihilated. Moreover, they undergo rare elastic scatterings with a background medium at rest, over a timescale $\tau_S \gg \Tilde{\tau}_F$. Working in the relaxation-time approximation (RTA), we have the following non-linear equation of motion:
\begin{equation}\label{nonlinearboltzmanncattaneo}
p^\mu \partial_\mu f =-\dfrac{u_\mu p^\mu}{\Tilde{\tau}_F} [f_{\text{eq}}(-\beta_\alpha p^\alpha)-f]+\dfrac{p^t}{\tau_S} [f_{\text{eq}}(\widetilde{\beta} p^t)-f] \qquad\qquad (\text{with }u^\mu \propto \beta^\mu),
\end{equation}
where $f_{\text{eq}}(z)=(e^z-1)^{-1}$ is the local-equilibrium occupation number.

The first collisional term models inter-particle collisions within the gas, which drive it towards local equilibrium. The second term models scatterings with the external medium, which force the gas to relax to a local-equilibrium state that is at rest relative to the medium.

The four-vector $\beta^\alpha$ depends on $f$ via Landau matching, whereas the scalar $\widetilde{\beta}$ is constructed so that the corresponding local-equilibrium distribution has the same energy density (as measured in the rest frame of the background medium) as $f$. To understand why this must be the case, we recall that the energy-momentum tensor of the gas is
\begin{equation}
T^{\mu \nu}=\int \dfrac{gd^3 p}{(2\pi)^3 p^t} p^\mu p^\nu f  \qquad \qquad (\text{with }g\text{ the spin degeneracy}) \, .
\end{equation}
Hence, multiplying both sides of \eqref{nonlinearboltzmanncattaneo} by $gp^\nu/p^t$, and integrating over all momenta, we obtain
\begin{equation}
\partial_\mu T^{\mu \nu}=-\dfrac{u_\mu}{\Tilde{\tau}_F} \int \dfrac{gd^3 p}{(2\pi)^3 p^t} p^\mu p^\nu [f_{\text{eq}}(-\beta_\alpha p^\alpha)-f] + \dfrac{1}{\tau_S} \int \dfrac{gd^3 p}{(2\pi)^3 } p^\nu [f_{\text{eq}}(\widetilde{\beta} p^t)-f] \, .
\end{equation}
Since collisions among the gas particles conserve total four-momentum, the first term on the right-hand side must vanish. Moreover, we assume that scatterings with the external medium are elastic, i.e. they conserve the energy measured in the medium's rest frame. Therefore, the time component of the second term in the rest frame of the external medium must vanish. This yields the matching conditions
\begin{equation}
\int \dfrac{gd^3 p}{(2\pi)^3 p^t} (-u_\mu p^\mu) p^\nu [f_{\text{eq}}(-\beta_\alpha p^\alpha)-f] =0 \, , \qquad \qquad \int \dfrac{gd^3 p}{(2\pi)^3 } p^t [f_{\text{eq}}(\widetilde{\beta} p^t)-f] = 0 \, .
\end{equation}

If we linearize the above equations around a global equilibrium state $f_{\text{eq}}(p^t/T)\equiv f_{\text{eq}}$ (with $T=\text{const}$), we obtain the system
\begin{equation}\label{linearRTA}
\begin{split}
& (\partial_t +n^j \partial_j) \delta f =\Tilde{\tau}_F^{-1} [ f_{\text{eq}}(1{+}f_{\text{eq}})p^\alpha \delta\beta_\alpha-\delta f]+\tau_S^{-1} [- f_{\text{eq}}(1{+}f_{\text{eq}})p^t \delta\widetilde{\beta}-\delta f] \, , \\
& \delta\beta_t = \dfrac{\displaystyle\int d^3 p \, p^t \delta f}{\displaystyle\int d^3 p \, f_{\text{eq}}(1{+}f_{\text{eq}})(p^t)^2 } =-\delta \widetilde{\beta}\, , \\
& \delta\beta_j = \dfrac{\displaystyle\int d^3 p \, p^j \delta f}{\displaystyle\int d^3p \, f_{\text{eq}}(1{+}f_{\text{eq}})(p^j)^2} \qquad (\text{no sum over } j) \, . \
\end{split}
\end{equation}

In the following, this will be regarded as our UV theory.

\subsection*{\textit{Step 2:} Recast the UV theory in standard self-adjoint form}
\vspace{-0.3cm}

The quadratic perturbation to the free-energy density due to a kinetic fluctuation $\delta f$ is given by
\begin{equation}
2\Delta \mathcal{F}=\int \dfrac{gd^3 p}{(2\pi)^3} \dfrac{(\delta f)^2}{f_{\text{eq}}(1{+}f_{\text{eq}})\beta} \, .
\end{equation}
This suggests that we introduce the rescaled degree of freedom
\begin{equation}
\Psi = \sqrt{\dfrac{g}{(2\pi)^3f_{\text{eq}}(1{+}f_{\text{eq}})\beta}} \, \delta f \, .
\end{equation}
In this way, the Onsager inner product is just $(\Psi,\Phi)=\int d^3 p \Psi^* \Phi$, and thus $\mathcal{H}=L^2(\mathbb{R}^3)$.
Hence, defined the fast timescale $\tau_F^{-1}\equiv\Tilde{\tau}_F^{-1}+\tau_S^{-1}$ (which in the EFT limit is almost indistinguishable from $\Tilde{\tau}_F^{-1}$), in conformity with the main text definition, we can rewrite the system \eqref{linearRTA} as follows
\begin{equation}\label{linearRTARescaled}
\begin{split}
& (\partial_t +n^j \partial_j) \Psi=(\tau_F^{-1}-\tau_S^{-1}) \sqrt{\dfrac{g f_{\text{eq}}(1{+}f_{\text{eq}})}{(2\pi)^3\beta}} p^j \delta\beta_j +  \tau_F^{-1} \sqrt{\dfrac{g f_{\text{eq}}(1{+}f_{\text{eq}})}{(2\pi)^3\beta}} p^t  \delta\beta_t -\tau_F^{-1}\Psi  \, , \\
& \delta\beta_\alpha = \dfrac{\displaystyle\int d^3 p \, \sqrt{\dfrac{gf_{\text{eq}}(1{+}f_{\text{eq}})}{(2\pi)^3 \beta}} p^\alpha \Psi}{\displaystyle\int d^3p \, \sqrt{\dfrac{g f_{\text{eq}}(1{+}f_{\text{eq}})}{(2\pi)^3 \beta}}p^\alpha \sqrt{\dfrac{g f_{\text{eq}}(1{+}f_{\text{eq}})}{(2\pi)^3 \beta}}p^\alpha} \qquad (\text{no sum over } \alpha) \, . \\
\end{split}
\end{equation}
Finally, defined the orthonormal vectors
\begin{equation}
e^\alpha = \dfrac{\sqrt{\dfrac{g f_{\text{eq}}(1{+}f_{\text{eq}})}{(2\pi)^3\beta}} p^\alpha}{\bigg|\bigg|\sqrt{\dfrac{g f_{\text{eq}}(1{+}f_{\text{eq}})}{(2\pi)^3\beta}} p^\alpha\bigg|\bigg|} \qquad (\text{no sum over } \alpha) \, ,
\end{equation}
the first line of \eqref{linearRTARescaled} reduces to
\begin{equation}
(\partial_t +n^j \partial_j) \Psi=(\tau_F^{-1}-\tau_S^{-1}) \sum_j e^j (e^j,\Psi) +  \tau_F^{-1} e^t (e^t,\Psi) -\tau_F^{-1}\Psi \, ,   
\end{equation}
which is written in standard self-adjoint form. From this, we can immediately read off the relevant operators:
\begin{equation}
\s=\tau_F^{-1} \left[1-\sum_\alpha e^\alpha (e^\alpha,*)\right ]+\tau_S^{-1}\sum_j e^j (e^j,*) \, , \qquad \qquad \E^j =n^j \, .
\end{equation}

\vspace{-0.3cm}
\subsection*{\textit{Step 3:} Identify the slow and fast Hilbert subspaces}
\vspace{-0.3cm}

Every state admits an orthogonal decomposition $\Psi=\sum_\alpha \Psi_\alpha e^\alpha +\Psi^\perp$, with $\Psi_\alpha=(e^\alpha,\Psi)$, and $(e^\alpha,\Psi^\perp)=0$. This decomposition diagonalizes $\s$, since $\s e^t=0$, $\s e^j=\tau_S^{-1}e^j$, and $\s\Psi^\perp=\tau_F^{-1}\Psi^\perp$. Hence, we immediately conclude that $\mathcal{H}_S=\text{span}\{e^\alpha\}$ and $\mathbb{S}=\sum_\alpha e^\alpha (e^\alpha,*)$. From this, it immediately follows that $\Psi_S=\sum_\alpha \Psi_\alpha e^\alpha$ and $\Psi_F=\Psi^\perp$.

\subsection*{\textit{Step 4:} Zeroth-order theory}

Using the orthonormal set $\{e^t,e^x,e^y,e^z\}$ as a basis of $\mathcal{H}_S$, the compressions $\s_{\mathcal{H}_S}$ and $\E_{\mathcal{H}_S}^j$ are represented as the $4\times 4$ matrices with elements $(e^\alpha,\s e^\beta)$ and $(e^\alpha,\E^j e^\beta)$, respectively. This gives
\begin{equation}
\partial_t 
\begin{bmatrix}
\Psi_t \\
\Psi_x \\
\Psi_y \\
\Psi_z \\
\end{bmatrix}=-
\begin{bmatrix}
0 & c_S\partial_x & c_S\partial_y & c_S\partial_z \\
c_S\partial_x & \tau_S^{-1} & 0 & 0 \\
c_S\partial_y & 0 & \tau_S^{-1} & 0 \\
c_S\partial_z & 0 & 0 & \tau_S^{-1} \\
\end{bmatrix}
\begin{bmatrix}
\Psi_t \\
\Psi_x \\
\Psi_y \\
\Psi_z \\
\end{bmatrix}+\mathcal{O}(\tau_F) \qquad \qquad (\text{with }c_S=1/\sqrt{3}) \, ,
\end{equation}
which is just Cattaneo's theory of heat transfer (expressed using Onsager's natural variables). The fact that Cattaneo's speed of propagation turns out to be $1/\sqrt{3}$ reflects the massless nature of the gas: on timescales much shorter than $\tau_S$ but much longer than $\tau_F$, the bosons behave as an ultrarelativistic fluid in local equilibrium, with sound speed $1/\sqrt{3}$.

\subsection*{\textit{Step 5:} EFT corrections}

The matrix elements of the first-order EFT corrections are $(e^\alpha,\mathbb{S}\E^j\mathbb{F}\s_{\mathcal{H}_F}^{-1}\mathbb{F}\E^k \mathbb{S}e^\beta)$. By symmetry, the inner products $(e^t,\mathbb{S}\E^j\mathbb{F}\s_{\mathcal{H}_F}^{-1}\mathbb{F}\E^k \mathbb{S}e^l)$ and $(e^l,\mathbb{S}\E^j\mathbb{F}\s_{\mathcal{H}_F}^{-1}\mathbb{F}\E^k \mathbb{S}e^t)$ vanish. Moreover, we have that $\s_{\mathcal{H}_F}^{-1}=\tau_F$. Thus,
\begin{equation}
\begin{split}
& (e^t,\mathbb{S}\E^j\mathbb{F}\s_{\mathcal{H}_F}^{-1}\mathbb{F}\E^k \mathbb{S}e^t)=\tau_F (n^j e^t,\mathbb{F}n^ke^t)= 0 \qquad\qquad (\text{because }n^k e^t \propto e^k\in \mathcal{H}_S) \, , \\
& (e^l,\mathbb{S}\E^j\mathbb{F}\s_{\mathcal{H}_F}^{-1}\mathbb{F}\E^k \mathbb{S}e^h)=\tau_F (n^j e^l,\mathbb{F} n^k e^h)=\dfrac{\tau_F}{5} \left(\delta^{jk}\delta^{lh}+\delta^{jh}\delta^{lk}-\dfrac{2}{3}\delta^{jl}\delta^{kh}\right) \, . \\
\end{split}
\end{equation}
Hence, we have the following dynamics:
\begin{equation}
\begin{split}
& \partial_t \Psi_t +c_S \partial^j \Psi_j =\mathcal{O}(\tau_F^2) \, , \\
& \partial_t \Psi_l +\tau_S^{-1}\Psi_l+c_S \partial_l \Psi_t =\dfrac{\tau_F}{5} \partial^j  \left(\partial_j \Psi_l+\partial_l \Psi_j-\dfrac{2}{3}\delta_{jl}\partial^k \Psi_k\right) +\mathcal{O}(\tau_F^2) \, , \\
\end{split}
\end{equation}
from which we conclude that $\mathfrak{D}_1=0$, $\mathfrak{D}_2=1/5$, and $\mathfrak{D}_3=0$. As can be seen, the only first-order effect is the shear viscosity associated with $\Psi_l$, which is proportional to the flow velocity $\delta u_l$ of the gas. Indeed, the coefficient $\tau_F/5$ coincides with the shear diffusivity coefficient $\eta/(\varepsilon+P)$ modeled within RTA.

\newpage
\section{A causal and stable representative of the first-order EFT}
\vspace{-0.2cm}

The equation of motion
\begin{equation}\label{parabolic}
(\partial_t+\s_{\mathcal H_S}+{\E^j}_{\mathcal H_S}\partial_j)\Psi_S=\tau_F\mathbb{D}^{jk}\partial_j \partial_k\Psi_S+\mathcal{O}(\tau_F^2) \, ,
\end{equation}
truncated to first order,
is in general parabolic, and therefore acausal. However, a causal representative of the same EFT can be obtained through the perturbative field redefinition $\Psi_S=[1+\tau_F \mathbb{A}^\mu \partial_\mu/2 +\mathcal{O}(\tau_F^2)]\Tilde{\Psi}_S$, where $\mathbb{A}^\mu$ are four operators on $\mathcal{H}_S$. Inserting this change of variables into \eqref{parabolic}, and multiplying both sides by $1+\tau_F (\mathbb{A}^\mu)^\dagger \partial_\mu/2 +\mathcal{O}(\tau_F^2)$ to preserve self-adjointness, we obtain
\begin{equation}\label{Maybehyperbolic}
\begin{split}
&\bigg(1+\tau_F\dfrac{(\mathbb{A}^t)^\dagger \s_{\mathcal H_S}{+}\s_{\mathcal H_S}\mathbb{A}^t}{2}\bigg) \partial_t \Tilde{\Psi}_S+\s_{\mathcal H_S} \Tilde{\Psi}_S +\bigg({\E^j}_{\mathcal H_S}+\tau_F\dfrac{(\mathbb{A}^j)^\dagger \s_{\mathcal H_S}{+}\s_{\mathcal H_S}\mathbb{A}^j}{2}\bigg) \partial_j \Tilde{\Psi}_S \\
&=\tau_F\bigg[\bigg(\mathbb{D}^{jk}-\dfrac{(\mathbb{A}^k)^\dagger {\E^j}_{\mathcal H_S}{+}{\E^j}_{\mathcal H_S}\mathbb{A}^k}{2}\bigg)\partial_j \partial_k-\dfrac{(\mathbb{A}^j{+}{\E^j}_{\mathcal H_S}\mathbb{A}^t)^\dagger{+}\mathbb{A}^j{+}{\E^j}_{\mathcal H_S}\mathbb{A}^t}{2} \partial_j \partial_t-\dfrac{(\mathbb{A}^t)^\dagger{+}\mathbb{A}^t}{2} \partial_t^2\bigg]\Tilde{\Psi}_S+\mathcal{O}(\tau_F^2) \, .\\
\end{split}
\end{equation}
To cancel the $\partial_t \partial_j$ terms, let us set $\mathbb{A}^t=\alpha$ (with $\alpha$ a positive constant) and $\mathbb{A}^j=-\alpha {\E^j}_{\mathcal H_S}$. Then, we obtain
\begin{equation}\label{Hyperbolic}
\begin{split}
\left(1+\tau_F \alpha \s_{\mathcal H_S}\right) \partial_t \Tilde{\Psi}_S &+\s_{\mathcal H_S} \Tilde{\Psi}_S +\bigg({\E^j}_{\mathcal H_S}-\tau_F\alpha\dfrac{{\E^j}_{\mathcal H_S} \s_{\mathcal H_S}{+}\s_{\mathcal H_S}{\E^j}_{\mathcal H_S}}{2}\bigg) \partial_j \Tilde{\Psi}_S \\
&=\tau_F\big[\big(\mathbb{D}^{jk}{+}\alpha {\E^j}_{\mathcal H_S} {\E^k}_{\mathcal H_S}\big)\partial_j \partial_k-\alpha \partial_t^2\big]\Tilde{\Psi}_S+\mathcal{O}(\tau_F^2) \, .\\
\end{split}
\end{equation}
This equation is EFT-equivalent to \eqref{parabolic}. We will now prove that, if there is a speed $w<1$ such that $||n_j {\E^j}_{\mathcal H_S}||\leq w$ for all $n_j \in S^2$ (i.e. if the zeroth-order theory is strictly subluminal), and $\tau_F$ is sufficiently small, then one can always choose $\alpha$ such that \eqref{Hyperbolic} (truncated to first order) is causal and stable. Note that \eqref{Hyperbolic} introduces additional non-hydrodynamic modes into the dynamics, but these modes necessarily decay on a timescale $\tau_F$ \cite{GavassinoReallInitialData:2026xjw}.

\vspace{-0.2cm}
\subsection*{Causality}
\vspace{-0.2cm}

Up to overall constant factors, the characteristic polynomial is $\mathcal{P}(\xi_\mu)=\det\big[ \xi_t^2-\big(\alpha^{-1}\mathbb{D}^{jk}{+} {\E^j}_{\mathcal H_S} {\E^k}_{\mathcal H_S}\big)\xi_j \xi_k\big]$. Let $\xi_\mu=(\xi v,\xi n_j)$ be a root of $\mathcal{P}$ (with $\xi\in \mathbb{R}$, $v \in \mathbb{C}$, and $n_j \in S^2$). Then, there exists a normalized vector $\Phi_S\in \mathcal{H}_S$ such that $\big[ v^2-\big(\alpha^{-1}\mathbb{D}^{jk}{+} {\E^j}_{\mathcal H_S} {\E^k}_{\mathcal H_S}\big)n_j n_k\big]\Phi_S=0$. Taking the inner product with $\Phi_S$, we obtain
\begin{equation}
v^2=\alpha^{-1}(\Phi_S, \mathbb{D}^{jk}n_j n_k \Phi_S )+({\E^j}_{\mathcal H_S} n_j\Phi_S, {\E^k}_{\mathcal H_S} n_k \Phi_S )\, .
\end{equation}
Both terms on the right-hand side are non-negative, and thus $v^2\geq 0$ (i.e. all roots are real). Moreover, $v^2 \leq \alpha^{-1}+w^2$. Hence, by taking $\alpha$ large enough, we can always enforce $v^2\leq 1$, and thus causality.

\vspace{-0.2cm}
\subsection*{Stability}
\vspace{-0.2cm}

Since the equations are causal, it will be enough to prove stability of Fourier modes in the rest frame, as stability in all other frames will immediately follow \cite{GavassinoSuperlum2021}. To that end, let us set $\Tilde{\Psi}_S\propto e^{ik_j x^j -i\omega t}$, with $k_j=k n_j$, where $k \in \mathbb{R}$ and $n_j \in S^2$. Then, \eqref{Hyperbolic} (truncated to first order) becomes
\begin{equation}\label{HyperbolicFourier}
\begin{split}
&\tau_F\alpha (i\omega)^2\Tilde{\Psi}_S-\left(1+\tau_F \alpha \s_{\mathcal H_S}\right) i\omega \Tilde{\Psi}_S+\s_{\mathcal H_S} \Tilde{\Psi}_S \\ &+ik\bigg(n_j{\E^j}_{\mathcal H_S}-\tau_F\alpha\dfrac{n_j{\E^j}_{\mathcal H_S} \s_{\mathcal H_S}{+}\s_{\mathcal H_S}n_j{\E^j}_{\mathcal H_S}}{2}\bigg) \Tilde{\Psi}_S +\tau_F\big[n_j\mathbb{D}^{jk}n_k{+}\alpha (n_j{\E^j}_{\mathcal H_S})^2\big]k^2 \Tilde{\Psi}_S=0 \, .\\
\end{split}
\end{equation}
Taking the inner product with $\Tilde{\Psi}_S$, and assuming that it is normalized, we obtain a quadratic equation of the form $a(i\omega)^2-b(i\omega)+c_1+i c_2=0$, with
\begin{equation}\label{abc1c2Estimates}
\begin{split}
a ={}& \tau_F \alpha \, , \\
b={}& \langle 1+\tau_F \alpha \s_{\mathcal H_S} \rangle \, , \\
c_1 ={}& \langle\s_{\mathcal H_S}+\tau_F\big[n_j\mathbb{D}^{jk}n_k{+}\alpha (n_j{\E^j}_{\mathcal H_S})^2\big]k^2\rangle \, , \\
c_2 ={}& k \bigg\langle n_j{\E^j}_{\mathcal H_S}-\tau_F\alpha\dfrac{n_j{\E^j}_{\mathcal H_S} \s_{\mathcal H_S}{+}\s_{\mathcal H_S}n_j{\E^j}_{\mathcal H_S}}{2}\bigg\rangle  \, ,\\
\end{split}
\end{equation}
where we have introduced the shorthand notation $\langle * \rangle \equiv(\Tilde{\Psi}_S,\, * \, \Tilde{\Psi}_S)$. Solving the quadratic, we find
\begin{equation}
i\omega = \dfrac{b\pm \sqrt{b^2-4a(c_1+ic_2)}}{2a} \, .
\end{equation}
\newpage
Stability requires that $\mathfrak{Re}(i\omega)_\pm\geq 0$. Since $a,b>0$, we only need to verify that $|\mathfrak{Re}\sqrt{b^2-4a(c_1+ic_2)}| \leq b$.
Using the identity $|\mathfrak{Re}\sqrt{z}|\,{=}\,\sqrt{(|z|{+}\mathfrak{Re}\,z)/2}$, the above condition is found to be equivalent to the requirement that the quantity $b^2 c_1{-}ac_2^2$ be non-negative. Explicitly, we have
\begin{equation}\label{ugly}
b^2 c_1{-}ac_2^2= \langle 1{+}\tau_F \alpha \s_{\mathcal H_S} \rangle^2 \langle\s_{\mathcal H_S}+\tau_F k^2\big[n_j\mathbb{D}^{jk}n_k{+}\alpha (n_j{\E^j}_{\mathcal H_S})^2\big]\rangle-\tau_F \alpha  k^2 \bigg\langle n_j{\E^j}_{\mathcal H_S}{-}\tau_F\alpha\dfrac{n_j{\E^j}_{\mathcal H_S} \s_{\mathcal H_S}{+}\s_{\mathcal H_S}n_j{\E^j}_{\mathcal H_S}}{2}\bigg\rangle^2 \, .
\end{equation}
Now, let us observe that
\begin{equation}
\begin{split}
\bigg\langle n_j{\E^j}_{\mathcal H_S}{-}\tau_F\alpha\dfrac{n_j{\E^j}_{\mathcal H_S} \s_{\mathcal H_S}{+}\s_{\mathcal H_S}n_j{\E^j}_{\mathcal H_S}}{2}\bigg\rangle^2={}& \bigg\langle \dfrac{n_j{\E^j}_{\mathcal H_S}(1{-}\tau_F\alpha \s_{\mathcal H_S}){+}(1{-}\tau_F\alpha\s_{\mathcal H_S})n_j{\E^j}_{\mathcal H_S}}{2}\bigg\rangle^2 \\
\leq {}& \langle (n_j{\E^j}_{\mathcal H_S})^2\rangle \langle(1{-}\tau_F\alpha \s_{\mathcal H_S})^2\rangle\, ,\\
\end{split}
\end{equation}
where we have invoked the Cauchy-Schwarz inequality. Hence, \eqref{ugly} gives
\begin{equation}\label{ugly2}
\begin{split}
b^2 c_1{-}ac_2^2\geq {}& \tau_F \alpha k^2 \langle  (n_j{\E^j}_{\mathcal H_S})^2\rangle \left[ \langle 1{+}\tau_F \alpha \s_{\mathcal H_S} \rangle^2- \langle(1{-}\tau_F\alpha \s_{\mathcal H_S})^2\rangle\right]  \\
= {}& 4\tau_F^2 \alpha^2 k^2 \langle  (n_j{\E^j}_{\mathcal H_S})^2\rangle \left[ \langle\s_{\mathcal H_S} \rangle+\dfrac{\tau_F \alpha}{4} (\langle \s_{\mathcal H_S} \rangle^2- \langle\s_{\mathcal H_S}^2\rangle)\right]  \\
\geq {}& 4\tau_F^2 \alpha^2 k^2 \langle  (n_j{\E^j}_{\mathcal H_S})^2\rangle \langle\s_{\mathcal H_S} \rangle\left[ 1-\dfrac{\alpha\tau_F}{4 \tau_S} \right] \, , \\
\end{split}
\end{equation}
where we have used the inequality $ \langle\s_{\mathcal H_S}^2\rangle\leq \tau_S^{-1}  \langle\s_{\mathcal H_S}\rangle$, which follows from $\text{Spectrum}(\s_{\mathcal H_S})\subset [0,\tau_S^{-1}]$.

In conclusion, $\alpha\tau_F <4 \tau_S$ is a sufficient condition for the theory to be stable.

\label{lastpage}
\end{document}